\documentclass[pra,aps,twocolumn,eqsecnum,superscriptaddress,showpacs]{revtex4}

\usepackage{amsmath}
\usepackage{graphicx}
\usepackage{color}
\usepackage{hyperref}
\usepackage{ulem}

\begin{document}

\title{Light scattering from an atomic gas under conditions of quantum degeneracy}

\author{V.M. Porozova}
\affiliation{Department of Theoretical Physics, St.-Petersburg State Polytechnic University, 195251, St.-Petersburg, Russia}
\author{L.V. Gerasimov}
\affiliation{Faculty of Physics, M.V. Lomonosov Moscow State University, Leninskiye Gory 1-2, 119991, Moscow, Russia}
\author{M.D. Havey}
\affiliation{Department of Physics, Old Dominion University, Norfolk, VA 23529, United States}
\author{D.V. Kupriyanov}\email{kupr@dk11578.spb.edu}
\affiliation{Department of Theoretical Physics, St.-Petersburg State Polytechnic University, 195251, St.-Petersburg, Russia}

\date{\today}

\begin{abstract}
We consider a quantum theory of elastic light scattering from a macroscopic atomic sample existing in the Bose-Einstein condensate (BEC) phase. The dynamics of the optical excitation induced by an incident photon is influenced by the presence of incoherent scattering channels. For a sample of sufficient length the excitation transports as a polariton wave and the propagation Green's function obeys the scattering equation which we derive. The polariton dynamics could be tracked in the outgoing channel of the scattered photon as we show via numerical solution of the scattering equation for one-dimensional geometry. The results are analyzed and compared with predictions of the conventional macroscopic Maxwell theory for light scattering from a non-degenerate atomic sample of the same density and size.
\end{abstract}

\pacs{42.50.Ct, 42.50.Nn, 42.50.Gy, 34.50.Rk}


\maketitle

\section{Introduction}\label{Section_I}

\noindent Light scattering from ultracold atomic systems existing under conditions of quantum degeneracy is a challenging and intriguing issue for both quantum optics and atomic physics. Together, investigation of these combined fields is practically important for developing various quantum interface protocols between light and matter subsystems. Although light scattering from either degenerate Bose or Fermi gases is of strong interest, we consider in the current context the degenerate Bose gas only, which is most typical for alkali-metal systems. The superposed light and matter wave propagating as a single quantum optical excitation through a Bose-Einstein condensate (BEC) phase had been predicted in \cite{Politzer} even before BEC had been created in the laboratory. Since the first successful experimental realizations of BEC in alkali-metal systems reported in \cite{Cornell,Ketterle}, evident signatures of cooperative dynamics in light scattering from the condensate have been observed in a series of experiments. These include manifestation of superradiant behavior of Rayleigh scattering in  \cite{Pritchard,Schneble,Hilliard}, formation of superfluid vortexes induced by coherent optical processes in \cite{Cornell99,Dalibard00,Phillips06} and spin vortexes in \cite{Brachmann11}, and optical control of the BEC phase transition with Faraday imaging technique in \cite{Sherman16}. The strong coherent coupling of light with a sample led to the condensate fragmentation \cite{Pritchard,Schneble,Hilliard} and explanation of such a quite non-trivial optomechanical effect has been attempted in \cite{Schneble} in terms of a Kapitza-Dirac diffraction phenomenon.

The above experiments have encouraged development of theoretical insights towards deeper understanding and precise description of light scattering under conditions of quantum degeneracy and from BEC in particular. The basic concept of a master equation for the order parameter suggests a relevant approach based on time dependent generalization of the non-linear Schr\"{o}dinger (Gross-Pitaevskii) equation \cite{Wright97,WillHoll99,Dowling05}. The coherent effects of conversion of either linear or angular momentum from light to the condensate are associated with a stimulated Raman process mediating the dynamics of the order parameter \cite{Phillips06,Dowling05}. The superadiant properties of the Rayleigh scattering, observed in a BEC, was explained by making use of the effective Hamiltonian approach via the mechanism of cooperative emission induced by a coherent classical pump in \cite{MooreMeystre99,ZobayNikol06,Trifonov,Avetisyan13,Avetisyan15}.

In the present report we are focusing on a microscopic quantum theory of a single photon scattering towards ab-initio description of elastic light scattering from a macroscopic atomic sample existing in the quantum degenerate BEC phase. Following the second quantized formalism, Bogolubov theory \cite{Bogoliubov47} and Gross-Pitaevskii model \cite{Gross,Pitaevskii} we introduce a set of coupled and closed diagram equations for the polariton propagator contributing to the $T$-matrix and scattering amplitude. Our approach allows us to follow important density corrections to the quasi-energy structure caused by static interaction and radiation losses associated with the incoherent scattering. We are aiming to test validity of the conventional macroscopic Maxwell description for the quantum degenerate gas as well as to follow possible deviations with light scattering from a non-degenerate atomic sample of the same density and size.

This paper is organized as follows. In Section \ref{Section_II} we develop our general theoretical framework of light scattering from a quantum degenerate atomic gas. In Section \ref{Section_III} we derive the basic scattering equation via the Feynman diagram method (briefly explained in Appendix \ref{Appendix_A}) and discuss general properties of the Green's function (polariton propagator) responsible for transporting an optical excitation in a BEC sample. In Section \ref{Section_IV} we present the results of our numerical simulations for light scattering in a one-dimensional geometry; the calculational scheme is detailed in Appendix \ref{Appendix_B}. In Section \ref{Section_V} we make some concluding remarks.

\section{The scattering problem under conditions of quantum degeneracy}\label{Section_II}
The quantum-posed description of the photon scattering problem is based on the formalism of the $T$-matrix, which is defined by
\begin{equation}
\hat T(E)=\hat V+\hat V\frac{1}{E-\hat H}\hat V%
\label{2.1}%
\end{equation}
where $\hat H$ is the total system Hamiltonian consisting of the unperturbed part $\hat H_0$ and an interaction term $\hat V$ such that $\hat H=\hat H_0+\hat V$. The energy argument $E$ is an arbitrary complex parameter in Eq.~(\ref{2.1}), which approaches a real physical value upon constructing the scattering amplitude. The scattering process, evolving from an initial state $|i\rangle$ to the final state $|f\rangle$, is expressed by the following relation between the differential cross section and the scattering amplitude, given by the relevant $T$-matrix element considered as a function of the initial energy $E_i$:
\begin{equation}
{\,d\sigma}_{i\to f}=\frac{{\cal V}^2{\omega'}^2}{\hbar^2 c^4{(2\pi)}^2}{|T_{g'\mathbf e'\mathbf k';\,g\mathbf e\mathbf k}(E_i+i\,0)|}^2\,d\Omega%
\label{2.2}%
\end{equation}
Here the initial state $|i\rangle=|g;\mathbf{e},\mathbf{k}\rangle$ is specified by the incoming photon's wave vector $\mathbf k$, frequency $\omega\equiv \omega_k = ck$, and polarization vector $\mathbf e$, and the atomic system populates a particular collective ground state $|g\rangle$. In our case  $|g\rangle=|\mathrm{BEC}\rangle^N$ initially performs a collective state of $N$ atoms in the BEC phase.  The final state $|f\rangle=|g';\mathbf{e}',\mathbf{k}'\rangle$ is specified by a similar set of quantum numbers, excepting that $|g'\rangle$ can be now a disturbed condensate state for inelastic channels, and the solid angle $\Omega$ is directed along the wave vector of the outgoing photon $\mathbf k'$. The presence of a quantization volume $\cal V$ in this expression is caused by the second quantized structure of the interaction operators. The optical theorem links the total cross section with the diagonal $T$-matrix element
\begin{equation}
\sigma_{\mathrm{tot}}=-\frac{2{\cal V}}{\hbar c}\,\mathrm{Im}\, T_{ii}(E_i+i\,0)%
\label{2.3}%
\end{equation}
which gives a convenient tool for the cross-section evaluation via calculation of only one $T$-matrix element for the elastic forward scattering.

In the second quantized representation the interaction term $\hat V$ in Eq.~(\ref{2.1}), taken in the dipole long wavelength approximation \cite{ChTnDpRcGr,BerstLifshPitvsk,Kupriyanov17}, is given by
\begin{equation}
\hat V=-\sum\limits_n\int\,d^3r\left[ d^{\mu}_{nm}\hat{E}_{\mu}({\mathbf r})\hat{\Psi}^{\dag}_n(\textbf{r})\hat{\Psi}_m(\textbf{r})+h.c.\right]%
\label{2.4}%
\end{equation}
where $d^{\mu}_{nm}$ is the matrix element of the $\mu$-th vector component of an atomic dipole moment, where $n$ and $m$ respectively specify the excited and ground states of the atom. $\hat{E}_{\mu}({\mathbf r})$ is the $\mu$-th vector component of the electric field operator and for sake of generality we use co/contravariant notation for the vector and tensor indices. The operators $\hat{\Psi}_m(\textbf{r})$ and $\hat{\Psi}_n^\dagger(\textbf{r})$ are the second quantized annihilation and creation operators of an atom at position $\mathbf{r}$ respectively in the ground and excited states.  We will further consider a BEC consisting of the simplest two-level atoms with a ${}^{1}S_0$ ground state and ${}^{1}P_1$ excited state such that quantum numbers $n=0,\pm 1$ and $m=0$ respectively denote the single atom angular momentum projection of the excited and the ground states.

In accordance with the general concept of quantum degeneracy for the system ground state existing in the BEC phase at zero-temperature, see \cite{BerstLifshPitvsk}, we accept
\begin{equation}
\hat\Psi_0(\textbf{r}){|\mathrm{BEC}\rangle }^N=\Xi(\textbf{r}){|\mathrm{BEC}\rangle }^{N-1}%
\label{2.5}%
\end{equation}
where $\Xi(\textbf{r})$ is the order parameter (often termed the "wavefunction") of the condensate. We consider the BEC as a macroscopic object such that the order parameter is insensitive to any small variation of the number of particles in the condensate. Then the scattering amplitude, expressed by "on-shell" $T$-matrix elements contributing to Eqs.~(\ref{2.2}) and (\ref{2.3}) for the scattering of an incident photon of frequency $\omega$ to the outgoing photon of frequency $\omega'$, is given by
\begin{eqnarray}
\lefteqn{T_{fi}(E)=\frac{2\pi\hbar(\omega'\omega)^{1/2}}{{\cal V}}\iint\,d^3r'\,d^3r\sum\limits_{n',n}}%
\nonumber\\%
&\times&{(\mathbf{d}\cdot\mathbf{e}')}^*_{n'0}{(\mathbf{d}\cdot\mathbf{e})}_{0n}\,{\mathrm e}^{-i\mathbf{k}'\mathbf{r}'+i\mathbf{k}\mathbf{r}}\ \Xi^*(\mathbf r')\,\Xi(\mathbf r)%
\nonumber\\
\nonumber\\
&\times&\left(-\frac{i}{\hbar}\right)\int_0^{\infty}dt\;{\mathbf e}^{\frac{i}{\hbar}(E-E_0^{N-1}+i 0)t}\;i\,G_{n'n}(\mathbf{r}',t;\mathbf{r},0)%
\nonumber\\%
\label{2.6}%
\end{eqnarray}
where $E_0^{N-1}$ is the initial energy of the condensate consisting of $N-1$ particles. The internal dynamics of the scattering process is described by a single optical excitation evolving in the condensate
\begin{equation}
i G_{n'n}(\mathbf{r}',t';\mathbf{r},t)=\langle\mathrm{BEC}|T\Psi_{n'}(\mathbf r';t')\Psi_n^{\dag}(\mathbf r;t)|\mathrm{BEC}\rangle^{N-1}%
\label{2.7}%
\end{equation}
with projection onto the product of condensate and field vacuum states such that entirely
\begin{equation}
|\mathrm{BEC}\rangle^{N-1}\equiv|\mathrm{BEC}\rangle^{N-1}_{\mathrm{Atoms}}\times|0\rangle_{\mathrm{Field}}
\label{2.8}
\end{equation}
Eq.~(\ref{2.7}) defines the time ordered (causal) Green's function (propagator) associated with the polariton-type quasi-particle excitation superposed between field and atom and propagating through the condensate consisting of $N-1$ particles. The operators contributing to the polariton propagator are the original atomic operators transformed in the Heisenberg representation and dressed by the interaction process. In the matrix element of the $T$-matrix in the form (\ref{2.6}) the outer operators $\hat{V}$ in its basic definition (\ref{2.1}) are disclosed in the rotating wave approximation (RWA). Such an assumption is surely valid as far as we are interested in near resonant scattering when both the frequencies $\omega$ and $\omega'$ are close to the frequency of the atomic transition $\omega_0$.

The Green's function (\ref{2.7}), rewritten in the interaction representation, can be expanded in the perturbation theory series
\begin{eqnarray}
\lefteqn{i G_{n'n}(\mathbf{r}',t';\mathbf{r},t)}%
\nonumber\\%
&&=\langle\mathrm{BEC}|\hat{S}^{-1}T\left[\Psi^{(0)}_{n'}(\mathbf r';t')\Psi^{(0)\dagger}_n(\mathbf r;t)\hat{S}\right]|\mathrm{BEC}\rangle^{N-1}%
\nonumber\\%
&&=\langle\mathrm{BEC}|T\left[\Psi^{(0)}_{n'}(\mathbf r';t')\Psi^{(0)\dagger}_n(\mathbf r;t)\hat{S}\right]|\mathrm{BEC}\rangle^{N-1}
\label{2.9}%
\end{eqnarray}
where in the interaction representation the $\Psi$-operators are superscripted by zero-indices. We consider the condensate itself as a stable system, which should not be modified by the interaction (\ref{2.4}) without its advanced perturbation by an incoming photon. This should be justified by the requirement that the evolution operator
\begin{equation}
\hat{S}=T\exp\left[-\frac{i}{\hbar}\int_{-\infty}^{\infty}\hat{V}^{(0)}(t)\,\mathrm{e}^{-0\cdot |t|}\right]%
\label{2.10}
\end{equation}
does not change the BEC state such that $\hat{S}|\mathrm{BEC}\rangle^{N-1}=|\mathrm{BEC}\rangle^{N-1}$. Although this requirement seems as evidently accepted in assumptions of the RWA let us make an important remark concerning its applicability.

The condensate, considered as a physical object, is not an ideal gas. The small but physically important difference $E_0^N-E_0^{N-1}=\varepsilon_0\equiv\mu_c+E_0$ gives a binding energy for adding a particle into an atomic ensemble, which incorporates the chemical potential $\mu_c$ and the internal ground state energy $E_0$ of a single atom. The latter could be set as zero but in our derivation it is convenient to leave $E_0$ as a physical parameter. For the quantum degenerate gas, consisting of not extremely dense and weakly interacting atoms and fairly described in a framework of the Gross-Pitaevskii model \cite{Gross,Pitaevskii}, the following inequality is fulfilled
\begin{equation}
\mu_c\lesssim\frac{\hbar^2 k_0^2}{2 {\mathrm{m_A}}}\ll \hbar\gamma%
\label{2.11}%
\end{equation}
where $k_0\equiv\lambdabar_0^{-1}$ is wave number for a resonant photon, ${\mathrm{m_A}}$ is the atomic mass and $\gamma$ is the natural spontaneous decay rate for the upper state of the atom. In accordance with the model, see \cite{BerstLifshPitvsk}, the chemical potential for a homogeneous BEC is given by
\begin{equation}
\mu_c=n_0\!\int\! d^3r\;U(r)>0%
\label{2.12}%
\end{equation}
where $U(r)$ is an interaction potential in the system of two atoms and $n_0$ is the atomic density. The subtle point is that the interaction $U(r)$ incorporates both the short range repulsive part and the long range attractive dipole-dipole polarization interactions. The latter is also known as the Van-der-Waals interaction and  the related asymptotic behavior of the potential $U(r)$ is constructed in the second order of the same Hamiltonian (\ref{2.4}) but with keeping the terms beyond and alternative to the RWA concept.

The conflicting situation with double accounting of the interaction Hamiltonian (\ref{2.4}) can be resolved once we pay attention that the Van-der-Waals interaction is meaningful on a distance of an atomic scale $r\sim O(1)a_0$, where $a_0$ is the Bohr radius, but the optical coupling experiences the distances $r\sim \lambdabar_0\gg a_0$ .  That means that there is no intersection in the diagram representation of $U(r)$ with those, which are induced by the evolution operator  (\ref{2.10}), and which couples a pair of distant atoms where one is always excited.  In this case the evolution operator(\ref{2.9}) indeed does not affect the condensate state and the second line in Eq.~(\ref{2.9}) is valid beyond the restrictions of the RWA approach as far as the internal interaction in the atomic ensemble is weak and can be safely separated from the optical excitation dynamics mediated by the scattering process.

Inequality (\ref{2.11}) provides us the chemical potential as the smallest parameter of the theory and is fulfilled up to the densities $n_0\lambdabar_0^3\gtrsim 1$. This is a typical condition with considering a condensate consisting of alkali-metal atoms. From the physical point of view that means that we consider the BEC in conditions close to an ideal gas and assume the matrix elements in (\ref{2.4}), as well as the atomic energy structure in the perturbation theory expansion, the same as for independent atoms. Nevertheless we do not ignore the gas non-ideality and the interatomic interaction $U(r)$ in the ground state as far as it is crucially important for proper description of the general behavior of the order parameter $\Xi=\Xi(\mathbf{r},t)$ under the framework of the Gross-Pitaevskii model with including superfluidity as the main macroscopic quantum property of the condensate. In its main approximations our consideration is applicable up to the bound of $\mu_c\lesssim \hbar\gamma$

\section{Dynamics of the optical excitation in the condensate}\label{Section_III}

\subsection{Diagrammatic representation}
\noindent The polariton propagator (\ref{2.9}) can be expanded in the perturbation theory series and the appearing terms can be regrouped with the Feynman diagram method. The basic elements and definitions are listed in Appendix \ref{Appendix_A}. As far as the considered interaction processes are primary developing in a near resonance conditions we follow the RWA approach with keeping leading expansion terms. Eventually the polariton propagator can be constructed as a dressed Green's function of an excited atom and obeys the following Dyson-type diagram equation
\begin{equation}
\scalebox{0.38}{\hspace{-0.5cm}\includegraphics*{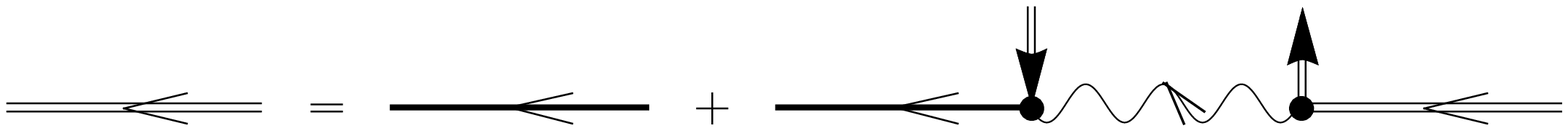}}
\label{3.1}
\end{equation}
where it is visualized as a doubly-straight line. This corresponds to the fact that the original atomic propagator is assumed as "dressed" here by all the interaction processes. The inward and outward vertical arrows image the order parameters and form the self-energy part responsible for coherent conversion of the excitation between the field, in which free dynamics is expressed by an undressed wavy line, and an atom subsequently recovered in the condensate phase. However, as consistent with this diagram equation, the above coherent process partly degrades because of interaction with the vacuum modes when the excited atom emits a photon spontaneously and escapes coherent dynamics with further drifting through the condensate as a spectator.

The latter process contributes in (\ref{3.1}) by an incomplete polariton propagator, which is imaged by a straight solid line in the diagrams and obeys the following Dyson-type equation
\begin{equation}
\scalebox{0.47}{\hspace{-0.5cm}\includegraphics*{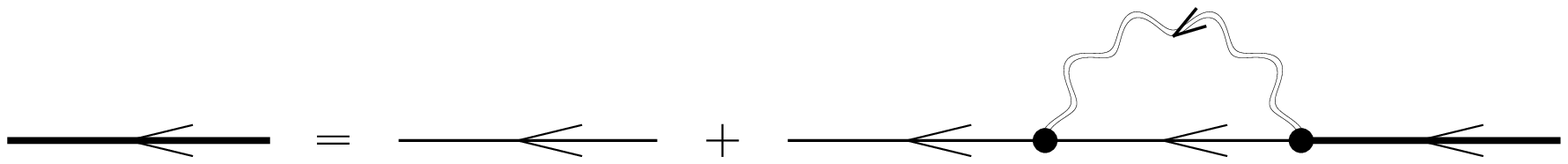}}
\label{3.2}
\end{equation}
which should be considered together with the equation for the dressed field propagator
\begin{equation}
\scalebox{0.47}{\hspace{-0.5cm}\includegraphics*{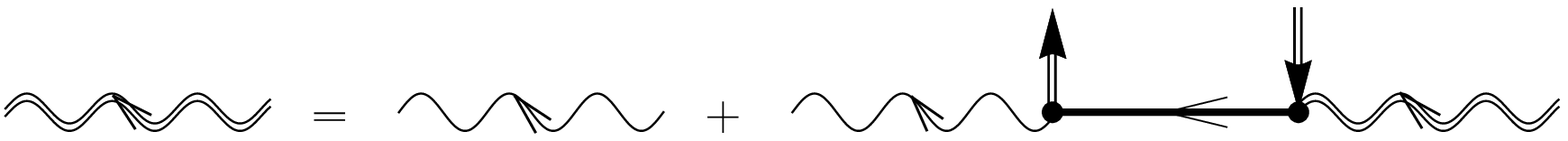}}
\label{3.3}
\end{equation}
These two diagram equations are closed with respect to each other and reproduce the self-consistent dynamics of an atomic dipole interacting with its environment similar to the conditions in a disordered atomic gas. Indeed any optical excitation created from the condensate has a chance to be incoherently re-emitted into the vacuum modes and transfer the atom, emitting such photon, out of the condensate phase. That is just described by the incomplete polariton propagator (\ref{3.2}) having a similar diagrammatic representation as of atomic excitation in a disordered gas. Such an incoherent scattering induces losses and leads to degradation of coherent dynamics supported by the self-energy operator in Eq.~(\ref{3.1}). In a natural optical association this process introduces the dielectric permittivity constructed similarly as in a disordered atomic gas of the same density.

\subsection{Incoherent losses and the dielectric permittivity of the condensate}

\noindent The self-energy part in (\ref{3.3}) (polarization operator) emphasizes the coherent structure of the matter state considered in conditions of quantum degeneracy. Nevertheless for an infinite, and locally homogeneous and isotropic medium, which physically requires that the sample size as well as inhomogeneity scale of the order parameter $\Xi(\mathbf{r})$ would be comparable or longer than the radiation wavelength, the solution of Eq.~(\ref{3.3}) is expected to be similar to the case of a disordered atomic gas of the same density. Indeed, both the vertices in the self-energy part of Eq.~(\ref{3.3}) are linked by the propagator (\ref{3.2}) in which the respective resonant excitation degrades on a time scale of natural decay when the excited atom can drift the distance much less than its radiation wavelength. Thus both the vertices are taken in proximal spatial points such that order parameter actually contributes to Eq.~(\ref{3.2}) as the local atomic density $n_0(\mathbf{r})=|\Xi(\mathbf{r})|^2$. With this simplification we can construct solution of equations (\ref{3.2}) and (\ref{3.3}) as for infinite, homogeneous and isotropic medium in closed analytical form and compare the result with similar performance of incoherent scattering process developing in disordered atomic gas.

\subsubsection{Analytical performance}
\noindent For sake of convenience and for further derivation we switch the primed and unprimed arguments and indices in notations of the Green's functions, see as an example Eq.~(\ref{2.9}), accordingly our definitions of Appendix \ref{Appendix_A} for the undressed functions. In stationary and homogeneous conditions these functions depend only on the difference between their spatial and time arguments. Then we can make a Fourier transform for the "dressed" Green's functions, constructed by the diagram method, and define
\begin{eqnarray}
\lefteqn{\hspace{-0.5cm}{\cal D}^{(E)}_{\mu\mu'}(\mathbf{k},\omega)}%
\nonumber\\%
&&\hspace{-0.5cm}=\!\int\!d^3\!R\int^{\infty}_{-\infty}\!\!d\tau\,%
\mathrm{e}^{i\omega\tau-i\mathbf{k}\cdot\mathbf{R}}\!\left.{\cal D}_{\mu\mu'}^{(E)}(\mathbf{R},\tau)\right|_{\scriptsize{\begin{array}{c}\mathbf{R}=\mathbf{r}-\mathbf{r}'\\ \tau=t-t'\end{array}}}%
\label{3.4}
\end{eqnarray}
for the photon propagator, fulfilling equation (\ref{3.3}), and
\begin{eqnarray}
\lefteqn{\hspace{-0.5cm}G^{(\gamma)}_{nn'}(\mathbf{p},E)}%
\nonumber\\%
&&\hspace{-0.5cm}=\!\int\!d^3\!R\int^{\infty}_{-\infty}\!\!d\tau\,%
\mathrm{e}^{\frac{i}{\hbar}E\tau-\frac{i}{\hbar}\mathbf{p}\cdot\mathbf{R}}\!\left.G_{nn'}^{(\gamma)}(\mathbf{R},\tau)\right|_{\scriptsize{\begin{array}{c}\mathbf{R}=\mathbf{r}-\mathbf{r}'\\ \tau=t-t'\end{array}}}%
\label{3.5}
\end{eqnarray}
for the incomplete polariton propagator, fulfilling equation (\ref{3.2}). The superscript $\gamma$ is added for associating such a propagator with excitation dynamics mediated by spontaneous scattering processes. In representation (\ref{3.4}) we assume the "dressed" positive frequency component of the vacuum Green's function (\ref{a.4}) with $\omega>0$ and the equivalence between the causal and retarded-type definitions for this case.

In the Fourier representation equation (\ref{3.3}) can be straightforwardly resolved with respect to the incomplete polariton propagator
\begin{eqnarray}
{\cal D}^{(E)}_{\mu\mu'}(\mathbf{k},\omega)&=&-\frac{4\pi\hbar\omega^2}{\omega^2_k-\epsilon(k,\omega)\omega^2}\left[\delta_{\mu\mu'}-c^2\,\frac{k_{\mu'}k_{\mu}}{\epsilon(k,\omega)\omega^2}\right]%
\nonumber\\%
\nonumber\\%
&\approx&-\frac{4\pi\hbar\omega^2}{\omega^2_k-\epsilon(\omega)\omega^2}\left[\delta_{\mu\mu'}-c^2\,\frac{k_{\mu'}k_{\mu}}{\epsilon(\omega)\omega^2}\right]%
\label{3.6}%
\end{eqnarray}
where
\begin{equation}
\epsilon(k,\omega)=1-\frac{4\pi}{\hbar}\,d^2_0\,n_0\,G^{(\gamma)}(\hbar k,\hbar\omega+\varepsilon_0)%
\label{3.7}%
\end{equation}
Here $d_0$ is the modulus of the transition dipole moment (the same for all the transitions), $n_0=|\Xi|^2=\mathrm{const}$ is the density of atoms, and for an isotropic medium with degenerate excited state ($E_n=\mathrm{const}_n$) we have
\begin{equation}
G^{(\gamma)}_{nn'}(\mathbf{p},E)\;=\;\delta_{nn'}\,G^{(\gamma)}(p,E)%
\label{3.8}%
\end{equation}
With taking into account inequality (\ref{2.11}) we expect negligible deviation in (\ref{3.7}) from the limit of immobile atoms and approximate $\epsilon(k,\omega)\approx\epsilon(0,\omega)\equiv\epsilon(\omega)$, which justifies the second line in Eq.~(\ref{3.6}). Equation (\ref{3.6}) (as well as similar tensor relations found later in the paper) is performed for Cartesian components $\mu,\mu'=x,y,z$, but for the case of spherical components $\mu,\mu'=0,\pm 1$ one has to change $\delta_{\mu\mu'}\to g_{\mu\mu'}=(-)^{\mu}\delta_{\mu,-\mu'}$.

The obtained result looks similar to that of a conventional medium beyond quantum degeneracy. As we can see, with reference to  \cite{BerstLifshPitvsk}, such a type  of "photon Green's function in a medium" can be associated with a fundamental solution of the macroscopic Maxwell equations where $\epsilon(\omega)$ is the dielectric permittivity of the medium. However in the case of quantum degeneracy both the excitations in the field and matter subsystems, i.e. photon and excited atom, transport through the sample in a superposed polariton mode, as suggested by the complete graph equation (\ref{3.1}). Although the association with a conventional medium is not intrinsically  consistent we shall call $\epsilon(\omega)$ as a dielectric permittivity of the condensate with having in mind in such analogy that it is constructed with involving only the contribution of over-condensate excitations created in the incoherent scattering process.

Equation (\ref{3.2}), decoded in the Fourier representation, contains the field Green's function (\ref{3.6}) contributing to the self-energy part in the form of the convolution integral with atomic propagator, see clarifying comment in Appendix \ref{Appendix_A}. As far as recovering of the incoherent losses as well as interaction with the quantized continuum are mostly important for near resonant conditions, we can expect that in the integral evaluation, the internal arguments are varied in sufficiently broad domains but located near $\omega\sim\omega_0$ and $k\sim k_0=\omega_0/c$, where $\omega_0=(E_n-E_0)/\hbar$ is the atomic transition frequency. Considering the field Green's function as an analytical function of detuning $\Delta=\omega-\omega_0$ in the complex half-plane where $\mathrm{Im}[\Delta]>0$ the integral over $\omega$ (approximated as integral over $\Delta$ in infinite limits) can be reliably reproduced by the residue at the pole point $\omega_E=(E-E_0)/\hbar$ (where $\Delta\to\Delta_E=(E-E_n)/\hbar$). In such an estimate we can safely ignore the small pole displacement associated with the Doppler shift as a negligible relativistic-type correction to the remaining integral evaluated over $\mathbf{k}$-variable.

In these assumptions, equation (\ref{3.2}) reads
\begin{equation}
\left[E-\frac{p^2}{2\mathrm{m_A}}-E_n-\Sigma^{(\gamma)}(p,E)\right]\,G^{(\gamma)}(p,E)=\hbar%
\label{3.9}%
\end{equation}
and the self-energy part $\Sigma^{(\gamma)}(p,E)$ is expressed by the sum
\begin{equation}
\Sigma^{(\gamma)}(p,E)=\Sigma^{(\mathrm{st})}(p,E)+\Sigma^{(\mathrm{rad})}(p,E)%
\label{3.10}%
\end{equation}
where the first term is given by
\begin{eqnarray}
\Sigma^{(\mathrm{st})}(p,E)&=&\frac{4\pi}{3}\int\frac{d^3k}{(2\pi)^3}\frac{d^2_0}{\epsilon\left(k,\omega_E\right)}%
\nonumber\\%
&\approx &\frac{4\pi}{3}\int\frac{d^3k}{(2\pi)^3}\frac{d^2_0}{\epsilon\left(\omega_E\right)}
\label{3.11}%
\end{eqnarray}
and can be associated with the interaction of the dipole with its own field in the environment of the over-condensate medium, created in the incoherent excitation process, see our comment above. The second term is given by
\begin{eqnarray}
\lefteqn{\hspace{-1.5cm}\Sigma^{(\mathrm{rad})}(p,E)=-\frac{8\pi}{3}d^2_0\int\frac{d^3k}{(2\pi)^3}\frac{\omega_E^2}{c^2k^2-\epsilon\left(k,\omega_E\right)\omega_E^2}}%
\nonumber\\%
\nonumber\\%
&&\hspace{-1cm}\approx -\frac{8\pi}{3}d^2_0\int\frac{d^3k}{(2\pi)^3}\frac{\omega_E^2}{c^2k^2-\epsilon\left(\omega_E\right)\omega_E^2}
\label{3.12}
\end{eqnarray}
and reveals radiation back action of the incoherent emission on the dipole's dynamics.

Equations (\ref{3.9})-(\ref{3.12}) and (\ref{3.7}) entirely construct one closed but quite complicated self-consistent equation for the incomplete propagator $G^{(\gamma)}(p,E)$, which has nonlinear and integral form. However the equation can be essentially simplified with applying faithful approximation, expressed by the second lines in Eqs.~(\ref{3.11}) and (\ref{3.12}), which assumes that in Eq.~(\ref{3.9}) the kinetic energy term for $p\sim\hbar k_0$ is small in comparison with the self-energy part. As we have pointed out above, this is justified by inequality (\ref{2.11}). In this approximation the dielectric permittivity $\epsilon(\omega)$ as well as the function $G^{(\gamma)}(p,E)$ (with $p\sim\hbar k_0$) can be found in analytical form once we resolve the problem with divergencies existing in both the contributions to the self-energy part (\ref{3.10}).

\subsubsection{Renormalization of the self-energy divergences}

\noindent Let us express contribution (\ref{3.11}) in the following form
\begin{equation}
\Sigma^{(\mathrm{st})}(p,E)\sim - \mathbf{d}\cdot\mathbf{E^{(\mathrm{vac})}}(\mathbf{0})-\mathbf{d}\left[\mathbf{E^{(\mathrm{med})}}(\mathbf{0})-\mathbf{E^{(\mathrm{vac})}}(\mathbf{0})\right]%
\label{3.13}%
\end{equation}
where we assumed that an atomic dipole $\mathbf{d}$ is located at the origin of the coordinate frame and the diverging integral (\ref{3.11}) was converted to the dipole's infinite electric field $\mathbf{E^{(\mathrm{med})}}(\mathbf{0})$ in the medium with dielectric constant $\epsilon$. We also subtracted and added the same quantity existing in vacuum with $\epsilon=1$. The vacuum term means the dipole self-action i.e. an artificial object of the theory, which reveals incorrectness of the dipole gauge on the distances comparable with atomic scale. The infinite energy, associated with this term, should be incorporated into the physical energy of the excited atom as internal energy of the point-like dipole particle. Then the second term in Eq.~(\ref{3.13}) is a physical quantity showing how the dipole self-action is modified in the environment of other dipoles. One expects that the incoherent scattering is a locally cooperative process and the selected dipole is indistinguishable from other proximal dipoles responding the driving field of an exciting photon. Then in accordance with the arguments performed in Refs.~\cite{Kupriyanov17,OurPRA2009,Javanainen97} we can accept the standard Lorentz-Lorenz interpretation of the field and energy shift, associated with static interaction of a collection of proximal dipoles
\begin{eqnarray}
\mathbf{E^{(\mathrm{med})}}(\mathbf{0})-\mathbf{E^{(\mathrm{vac})}}(\mathbf{0})&\to&\frac{4\pi}{3}n_0\mathbf{d}%
\nonumber\\%
\Sigma^{(\mathrm{st})}(p,E)\to -\frac{4\pi}{3}n_0d_0^2%
\label{3.14}
\end{eqnarray}
where we substituted $\mathbf{d}^2\to d_0^2$.

The second contribution (\ref{3.12}) can be interpreted as an interaction with the quantized vacuum continuum manifestable via the radiation Lamb-shift and spontaneous decay rate. With taking $\epsilon=1$ the integral transforms to
\begin{eqnarray}
\lefteqn{\hspace{-4cm}\Sigma^{(\mathrm{vac})}(p,E)=-\left.\frac{8\pi}{3}d^2_0\int\frac{d^3k}{(2\pi)^3}\frac{\omega_E^2}{c^2k^2-\omega_E^2-i0}\right|_{E\sim E_n}}%
\nonumber\\%
\nonumber\\%
&&\hspace{-2cm}\Rightarrow\hbar\Delta_L-\frac{i\hbar}{2}\gamma%
\label{3.15}
\end{eqnarray}
where $\Delta_L\to\infty$ is the vacuum Lamb-shift further renormalized and incorporated into the atomic energy $E_n$, "dressed" by interaction with the vacuum modes. The regularized integral given by the difference of (\ref{3.12}) and (\ref{3.15})  becomes converging and reproducible by residues at its pole points. Eventually we arrive at the following renormalization of the radiation correction to the self-energy part
\begin{equation}
\Sigma^{(\mathrm{rad})}(p,E)\Rightarrow -\frac{i\hbar}{2}\sqrt{\epsilon\left(\omega_E\right)}\gamma%
\label{3.16}%
\end{equation}
which contains both the radiation damping and energy shift modified by the radiation coupling with the over-condensate environment.

\subsubsection{The incomplete propagator in closed form}

\noindent With substituting renormalized self-energy parts (\ref{3.14}) and (\ref{3.16}) into (\ref{3.10}) and (\ref{3.9}) and in accordance with our definition of the dielectric permittivity given by Eq.~(\ref{3.7}) with $k\to 0$ we obtain the following equation
\begin{equation}
\epsilon\left(\omega-\frac{\mu_c}{\hbar}\right)=\frac{\omega-\omega_0-\displaystyle\frac{8\pi}{3\hbar}n_0d_0^2+\frac{i}{2}\sqrt{\epsilon(\omega)}\gamma}%
{\omega-\omega_0+\displaystyle\frac{4\pi}{3\hbar}n_0d_0^2+\frac{i}{2}\sqrt{\epsilon(\omega)}\gamma}
\label{3.17}%
\end{equation}
In the left-hand side the frequency argument of the permittivity is displaced by the  chemical potential $\mu_c$. This emphasizes the fact that for a single optical excitation from the condensate the extra action is needed, which is a meaningful part of binding energy $\varepsilon_0=E_0+\mu_c$ given by the chemical potential. Although in our model this displacement is rather small it recognizes a qualitatively important extension up to the case of a strongly non-ideal gas. But with neglecting it, we obtain an equation for the dielectric permittivity identical to an atomic ensemble consisting of cold disordered and randomly distributed atomic dipoles, see Ref.~\cite{OurPRA2009}.

Equation (\ref{3.17}) can be analytically solved and its solution can be applicable for the case of a inhomogeneous medium if the density $n_0=n_0(\mathbf{r})$ and order parameter $\Xi(\mathbf{r})$ are varied on a spatial scale comparable with the radiation wavelength or longer. Then equation (\ref{3.9}) suggests the following approximate form in the mixed space-frequency representation
\begin{eqnarray}
\lefteqn{\hspace{-1cm}\left[E+\frac{\hbar^2}{2\mathrm{m_A}}\triangle-E_n+\frac{4\pi}{3}\,n_0(\mathbf{r})\,d_0^2\right.}
\nonumber\\%
&&\hspace{-1cm}\left.+\frac{i\hbar}{2}\sqrt{\epsilon\left(\mathbf{r},\omega_E\right)}\;\gamma\right]
G^{(\gamma)}(\mathbf{r},\mathbf{r}';E)=\hbar\,\delta(\mathbf{r}-\mathbf{r}')%
\label{3.18}%
\end{eqnarray}
where we parametrized the dielectric constant $\epsilon=\epsilon(\mathbf{r},\omega)$ by its spatial dependence. Indeed, in this equation $G^{(\gamma)}(\mathbf{r},\mathbf{r}';E)$, considered as a function of $\mathbf{r}-\mathbf{r}'$, transports a single photon excitation, created from the immobile condensate, from point $\mathbf{r}'$ to point $\mathbf{r}$, which degrades on a spatial scale sufficiently less than $\lambdabar_0=k_0^{-1}$. Thus equation (\ref{3.18}) accepts only proximal spatial arguments $\mathbf{r}\sim\mathbf{r}'\sim(\mathbf{r}+\mathbf{r}')/2$ where $n_0=n_0(\mathbf{r})$ is approximately constant.

We have constructed the \textit{incomplete} polariton propagator (\ref{3.2}) in the form, which is similar to the \textit{complete} excited state propagator of a single atom in a disordered atomic gas of the same density. Such an analogy, emphasizing the similarity in spontaneous scattering from both the systems, was expectable and prefaced this part of our derivation. Nevertheless, as was pointed out above, the analogy is not so straightforward and in the conditions beyond the Gross-Pitaevskii model (i.e. for a non-ideal quantum gas with strong internal coupling) it could appear important deviations in description of such physically different systems.

\subsection{The complete polariton propagator}\label{Section_III_C}
\noindent With decoding the diagram equation (\ref{3.1}) for the complete polariton propagator we extend spontaneous dynamics, described by  Eq.~(\ref{3.18}), with involving the process of coherent conversion of the excitation between field and condensate
\begin{eqnarray}
\lefteqn{\left[E+\frac{\hbar^2}{2\mathrm{m_A}}\triangle-E_n+\frac{4\pi}{3}\,n_0(\mathbf{r})\,d_0^2\right.}
\nonumber\\%
&&\left.+\frac{i\hbar}{2}\sqrt{\epsilon\left(\mathbf{r},\omega_E\right)}\;\gamma\right]%
G_{nn'}(\mathbf{r},\mathbf{r}';E)%
\nonumber\\%
&&-\sum_{n''}\int\!\!d^3\!r''\;\Sigma_{nn''}^{(\mathrm{c})}(\mathbf{r},\mathbf{r}'';\!E)\;G_{n''n'}(\mathbf{r}''\!,\mathbf{r}';\!E)%
\nonumber\\%
\nonumber\\%
&&=\hbar\,\delta_{nn'}\delta(\mathbf{r}\!-\!\mathbf{r}')%
\label{3.19}%
\end{eqnarray}
The kernel of the respective integral self-energy operator (with simplifying argument superscripted from double prime to single prime) is given by
\begin{eqnarray}
\hspace{-1cm}\Sigma_{nn'}^{(\mathrm{c})}(\mathbf{r},\mathbf{r}';E)&=&\frac{1}{\hbar}\sum_{\mu\mu'}\,\Xi(\mathbf{r})\,\Xi^{\ast}(\mathbf{r}')\;d^{\mu}_{n0}d^{\mu'}_{0n'}%
\nonumber\\%
&\times&D_{\mu\mu'}^{(E)}\left(\mathbf{r}-\mathbf{r}',\omega_E-\frac{\mu_c}{\hbar}\right)%
\label{3.20}%
\end{eqnarray}
where the vacuum field Green's function, expressed by the wavy line in the diagram equation (\ref{3.1}) and defined by Eqs.~(\ref{a.1}) and (\ref{a.4}), contributes here in the mixed space-frequency representation
\begin{eqnarray}
\lefteqn{\hspace{-1.5cm}D^{(E)}_{\mu\mu'}(\mathbf{R};\omega)\!\!=\!\!-i\!\int^{\infty}_{-\infty}\!\!d\tau\,%
\mathrm{e}^{i\omega\tau}\!\left.\langle T E^{(0)}_{\mu}(\mathbf{r},t)\,E^{(0)}_{\mu'}(\mathbf{r}',t')\rangle\right|_{\scriptsize\begin{array}{c}\tau\!=\!t\!-\!t'\\ \mathbf{R}\!=\!\mathbf{r}\!-\!\mathbf{r}'\end{array}}}%
\nonumber\\%
&=&-\hbar\frac{|\omega|^3}{c^3}\left\{i\frac{2}{3}h^{(1)}_0\left(\frac{|\omega|}{c}R\right)\delta_{\mu\nu}\right.%
\nonumber\\%
&+&\left.\left[\frac{X_{\mu}X_{\mu'}}{R^2}-\frac{1}{3}\delta_{\mu\mu'}\right]%
ih^{(1)}_2\left(\frac{|\omega|}{c}R\right)\right\}%
\label{3.21}
\end{eqnarray}
Here the averaging is over the vacuum state and $h^{(1)}_L(\ldots)$ with $L=0,2$ are the spherical Hankel functions of the first kind.

The derived equation (\ref{3.19}) traces the dynamics of a single particle excitation in the condensate with the assumption that the order parameter, density distribution, dielectric permittivity, etc. have a smooth profile on a mesoscopic scale, similar to the conventional macroscopic Maxwell theory. It visualizes as a Schr\"{o}dinger-type equation for an excited atom propagating in space and modified by interacting with the environment. Here the kinetic energy term is actually responsible for negligible drift of the excitation during the decay time when the transferred momentum of the polariton is limited by the value of $\hbar k_0$ in its order of magnitude. Nevertheless the optical excitation itself can propagate through the sample with much faster speed with approaching to speed of light, which can be demonstrated via solution of equation (\ref{3.19}) in the limit of infinite and homogeneous medium.

For an infinite, homogeneous and isotropic medium the solution of Eq.~(\ref{3.19}) can be found in the reciprocal space as a linear combination of the transverse and the longitudinal components with respect to the momentum argument. In our further estimates in this subsection, with considering the internal binding energy in the condensate as weak, we will ignore the chemical potential of the condensate as a negligible quantity in comparison with the basic spectral parameters such as spontaneous decay rate and recoil energy, see (\ref{2.11}). Then the Fourier components of the complete polariton propagator can be expanded as follows
\begin{eqnarray}
G_{nn'}(\mathbf{p},E)&=&G_{\parallel}(p,E)\,\frac{p_np_{n'}}{p^2}
\nonumber\\%
&&+\ G_{\perp}(p,E)\!\left[\delta_{nn'}-\frac{p_np_{n'}}{p^2}\right]
\label{3.22}
\end{eqnarray}
where, in accordance with the selection rules for the dipole moment operators in Eq.~(\ref{3.20}), we associated the vector indices (in Cartesian frame) in the quasi-particle momentum $\mathbf{p}$ with the quantum numbers of the atomic excited state.

The longitudinal and transverse components of the polariton propagator are respectively given by
\begin{eqnarray}
G_{\parallel}(p,E)\!&=&\!\hbar\left[E-E_n-\frac{p^2}{2\mathrm{m_A}}-\frac{8\pi}{3}n_0\,d_0^2\right.
\nonumber\\%
&&\hspace{1.2cm}\left.+\ \frac{i\hbar}{2}\sqrt{\epsilon(\omega_E)}\;\gamma\right]^{-1}%
\nonumber\\%
\nonumber\\%
G_{\perp}(p,E)\!&=&\!\hbar\left[E-E_n-\frac{p^2}{2\mathrm{m_A}}+\frac{4\pi}{3}\,n_0\,d_0^2\right.%
\nonumber\\%
&&\hspace{-2cm}\left.+\frac{i\hbar}{2}\sqrt{\epsilon(\omega_E)}\;\gamma-\;\frac{4\pi\,n_0\,d_0^2\,\omega_E^2}{\left(\omega_E^2-c^2p^2/\hbar^2\right)}\right]^{-1}%
\label{3.23}
\end{eqnarray}
With approaching the point of atomic resonance $E\to E_n$ the optical excitation shows behavior associating with that of a non-condensed disordered atomic gas. The collective dipole polarization is driven by the propagating field and the environment of proximal dipoles induces the well known  static Lorentz-Lorentz red shift  $-4\pi\,n_0\,d_0^2/3$ from the atomic resonance as it contributes to the transverse part of the propagator. However unlike a disordered gas there is an extra static frequency shift, induced by the polarization interaction with the condensate background, which is given by the last term in the right-hand side of Eq.~(\ref{3.23}). Indeed, considering the quasi-particle as immobile with negligible momentum $p\ll\hbar\omega/c$ the dependence on $E$ vanishes and this part of the interaction also becomes static. In this limit the transverse component of the polariton propagator coincides with its longitudinal part, such that the excitation process becomes isotropic with positive static shift $8\pi\,n_0\,d_0^2/3$.

The spectral behavior of the polariton propagator in the form (\ref{3.22}), (\ref{3.23}) consists of two branches. One is an atom-type excitation near atomic resonance $E\sim E_n$, on which we have commented above. Another resonance exists in the transverse part of the polariton propagator and is located near the energy $E\sim E_0+c\,p$, which is a pole feature of the last term in the denominator of the transverse component $G_{\perp}(p,E)$. This resonance describes the optical excitation propagating through the sample with near speed of light and creates the photon-type polariton branch. The detail discussion of spectral behavior of the polariton modes in the infinite and homogeneous medium is performed in Ref.~\cite{Ezhova16}.

In general, with an inhomogeneous configuration with the order parameter of arbitrary profile, equation (\ref{3.19}) accepts only numerical solution. In the next section we present such a solution in a one-dimensional geometry and compare the results with predictions of conventional macroscopic Maxwell theory.

\section{Results}\label{Section_IV}
\begin{figure}[tp]
{$\scalebox{0.55}{\includegraphics*{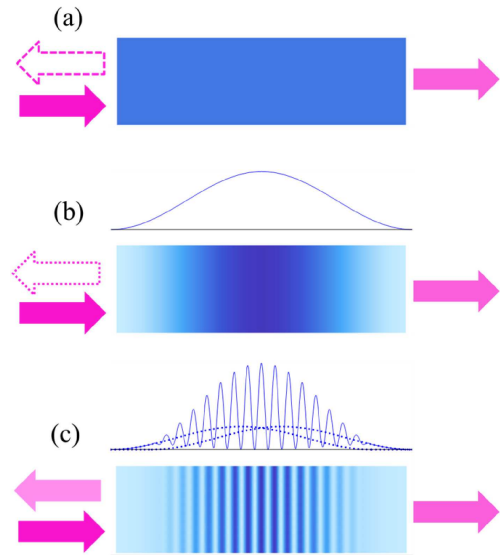}}$}
\caption{(Color online) Geometry of the considered one-dimensional scattering process: (a) a uniform BEC slab of depth $L$ and with the order parameter $\Xi=\sqrt{n_0}=\mathrm{const}_{z}$, where $n_0$ is the density of atoms; (b) an inhomogeneous distribution parametrized by the order parameter $\Xi(z)=\sqrt{n_0}\cos (\pi z/L)$; (c) interference of two matter waves created by the BEC sample (b) split in two fragments, see Eq.~(\ref{4.2}) and explanation in the text.}
\label{fig1}%
\end{figure}

\noindent Degenerate quantum gases have unique properties and are of particular interest in reduced spatial dimensionality \cite{Salomon,Levin}. This motivates us to consider initially our results for several instances of a one dimensional model. Further, equation (\ref{3.19}) is quite difficult for numerical solution in a general three-dimensional configuration. Below we perform results of our numerical simulations for a one-dimensional model expressed in terms of transmission and reflection of light from a slab atomic sample, where atoms can exist in either a quantum degenerate phase or as a disordered classical gas. The considered geometries are shown in Fig.~\ref{fig1} for three tested configurations - (a) a uniform slab of BEC with constant density, (b) an inhomogeneous distribution parameterized by the order parameter with a cosine profile, (c) interference of two matter waves for two BEC segments counter-propagating through each other. In the last case, as we show, such an internal motion of the overlapping condensate fragments can crucially modify the light scattering process.

\subsection{Smooth profile of the order parameter}
\noindent Any testable profile of the order parameter should be consistent with the physical model of the condensate and, in the case of weak internal coupling, performs as a possible solution of the Gross-Pitaevskii equation \cite{Pitaevskii,Gross}. In the macroscopic limit any homogeneous spatial profile of the order parameter can be suggested as an example of a Thomas-Fermi-type approximate solution, for which the shape can be fitted by varying the trapping potential. This approximation works for the condensate confined with an atomic trap where the period of free oscillation is longer than $2\pi/\mu_c$ with $\mu_c$ estimated (in a homogeneous limit) by Eq.~(\ref{2.12}) and it is based on priority of internal interaction. But even in the case of an ideal gas with $\mu_c\to 0$ the order parameter of a quite general profile can be accepted as well, but in this case as the ground state eigenfunction of the stationary single particle Schr\"{o}dinger equation in the trap potential.

As a first example, let us consider the case of a homogeneous degenerate quantum gas filling a slab of depth $L$ with the order parameter given by $\Xi=\sqrt{n_0}=\mathrm{const}_{z}$, which is shown in Fig.~\ref{fig1}(a). In a one-dimensional geometry, with applying the Fourier transform, the scattering equations (\ref{3.19}) can be rewritten as an infinite set of the algebraic equations, see Appendix \ref{Appendix_B} for derivation details. The obtained system of algebraic equations can be numerically solved, which give us the spectra of transmission $\mathrm{T}(\omega)$ and reflection $\mathrm{R}(\omega)$. The same quantities can be independently constructed via solution of the macroscopic Maxwell equations, see \cite{LaLfVIII}, and they are given by
\begin{eqnarray}
\mathrm{T}(\omega)&=&\left|\frac{2\sqrt{\epsilon(\omega)}}{2\sqrt{\epsilon(\omega)}\cos\psi(\omega)-i(1+\epsilon(\omega))\sin\psi(\omega)}\right|^2%
\nonumber\\%
\mathrm{R}(\omega)&=&\left|\frac{\sin\left[\psi(\omega)\right]}{\sin\left[\psi(\omega)-i\ln\frac{1-\sqrt{\epsilon(\omega)}}{1+\sqrt{\epsilon(\omega)}}\right]}\right|^2%
\label{4.1}%
\end{eqnarray}
where $\psi(\omega)=L\sqrt{\epsilon(\omega)}\omega/c$. With substituting here the dielectric permittivity (\ref{3.17}) (with canceled chemical potential) we arrive at the result predicted for a macroscopic disordered gas, see \cite{OurPRA2009}.

\begin{figure}[tp]
{$\scalebox{0.8}{\includegraphics*{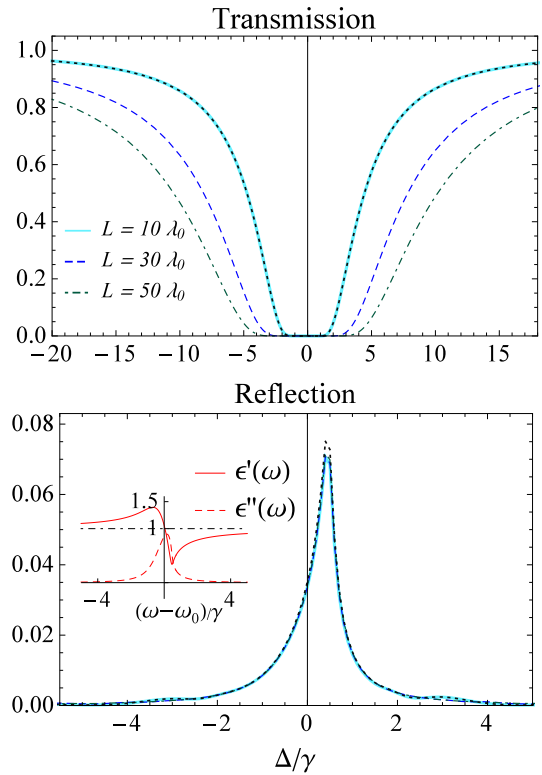}}$}
\caption{(Color online) The spectral dependences of transmission (upper panel) and reflection (lower panel) calculated as a solution of the scattering equations (\ref{3.19}) vs. comparative solution of the Maxwell equations (\ref{4.1}) in a one-dimenional geometry for a homogeneous medium with a slab geometry shown in Fig.~\ref{fig1}(a). The graphs are plotted as a function of detuming $\Delta=\omega-\tilde{\omega}_0$ from the displaced resonance frequency $\tilde{\omega}_0=\omega_0-\mu_c/\hbar$, see text. The results are performed for different sample depths $L$, scaled by the wavelength $\lambda_0$ at the atomic resonance, and for the density $n_0\lambdabar_0^3\sim 0.05$. The reflection spectra for different $L$ are unresolved in the graph with the plotted precision. The inset shows the dielectric permittivity of the sample $\epsilon(\omega)=\epsilon'(\omega)+i\epsilon''(\omega)$ given by solution of Eq.~(\ref{3.17}) as a function of $\omega-\omega_0$. Both the rounds of calculations give identical results, and to show this in the example of $L=10\lambda_0$ we additionally indicate (by dotted curve) the prediction of the macroscopic Maxwell theory.}
\label{fig2}%
\end{figure}

In Fig.~\ref{fig2} we compare the spectra of light transmission through and reflection from the condensate and disordered atomic gas of the same density $n_0\lambdabar_0^3\sim 0.05$ and in the geometry of Fig.~\ref{fig1}(a)The inset shows the dielectric permittivity given by solution of Eq.~(\ref{3.17}). Since an optical excitation from the condensate changes its energy the excitation spectrum of non-ideal degenerate quantum bosonic gas is red-shifted from atomic resonance by the value of the chemical potential. The shift is small and seems negligible as far as the condition (\ref{2.11}) is normally fulfilled for any dipole-type transition and in alkali-metal systems in particular. Thus we could safely ignore this shift with constructing the susceptibility for the condensate as the solution of Eq.~(\ref{3.17}). Nevertheless, we leave it in our reproduction of the spectral responses as far as such a red shift is a physical effect and can be visible in the transmission and reflection spectra. The red shift has been observed in the transmission spectrum of a BEC consisting of helium atoms on a spectrally narrow dipole forbidden magnetic-type transition \cite{NRV16}.

Surprisingly, but this global offset of the spectral profile is only one difference between the transmission and reflection spectra of degenerate and non-degenerate atomic gases. To demonstrate this we plotted the graphs as a function of detuning $\Delta=\omega-\tilde{\omega}_0$, where $\tilde{\omega}_0=\omega_0-\mu_c/\hbar$, and where we additionally displaced the spectra of a disordered gas on $\mu_c/\hbar$. We have obtained excellent, i.e. point by point, coincidence of degenerate and non-degenerate spectra despite the fact that they were calculated via solution of exceptionally different equations. The small deviation for reflection near its resonant point is a result of additional boundary contributions ignored in the Fourier transformation of the Laplace operator to the algebraic form of Eq.~(\ref{b.8}) and this incorrectness, as we have verified, softens in the macroscopic limit $L/\lambda_0\to\infty$. The reflection itself is weak but not negligible and results from the scattering from the sample edges and is enhanced by interference effect. Such an excellent coincidence of two independent rounds of calculations clearly indicates that for light scattering from an ensemble of atoms, with uniform density distribution, the optical response of the system is insensitive to either classical or quantum nature of statistical averaging.

\begin{figure}[tp]
{$\scalebox{0.8}{\includegraphics*{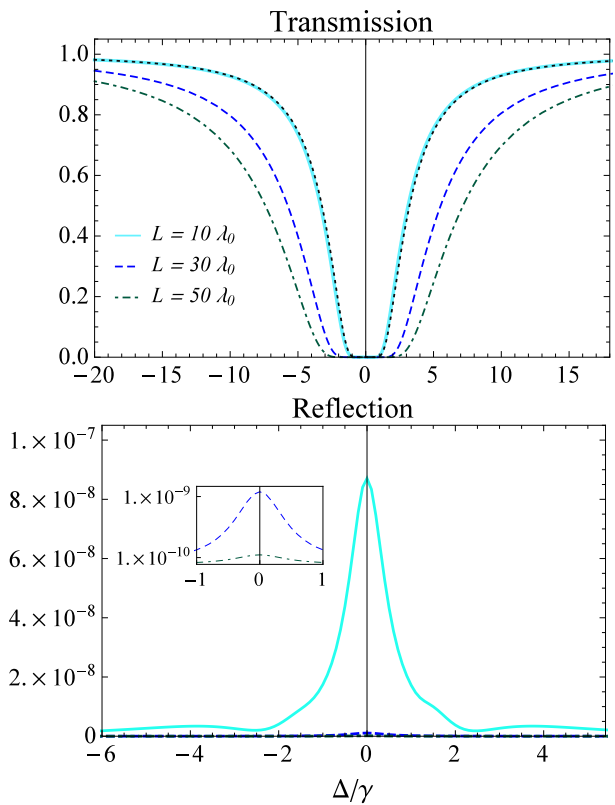}}$}
\caption{(Color online) Same as in Fig.~\ref{fig2} but for the density distribution parametrized by the order parameter $\Xi(z)=\sqrt{n_0}\cos (\pi z/L)$ for a slab geometry shown in Fig.~\ref{fig1}(b). In the case of a smooth sample boundary the backward scattering reveals a many orders of magnitude weaker process than in the case of sharp boundaries.}
\label{fig3}%
\end{figure}

This can be confirmed by similar calculations performed for the order parameter with a trigonometric profile $\Xi(z)=\sqrt{n_0}\cos (\pi z/L)$ ( in geometry of Fig.~\ref{fig1}(b)), and the results are shown in Fig.~\ref{fig3}. For this case we make additional simplifications with expanding $\sqrt{\epsilon\left(z,\omega\right)}$ in a Taylor series near the vacuum point $\epsilon=1$ and with keeping only the forwardly propagating wave in the macroscopic Maxwell description of the problem. Again the calculations show good (within the made approximations) agreement between both the approaches. We used the same peak density $n_0\lambdabar_0^3\sim 0.05$ and the same sample depths as in the plots of Fig.~\ref{fig2}. In the case of smoothed sample bounds with density profile $n_0(z)=n_0\cos^{2} (\pi z/L)$ the backward scattering is expected as many orders of magnitude weaker process because of vanishing boundary contributions. The latter can be seen via negligible response of the reflected light as follows from the calculation data shown in the lower panel of Fig.~\ref{fig3}.

\subsection{Interference of two counter-propagating BEC fragments}
\noindent Finally, let us consider the experimental configuration when, as a result of coherent interaction with light, a BEC sample is fractured into a number of macroscopic pieces \cite{Schneble,Hilliard}. To simplify the complicated experimental picture we model the process by the presence of only two fragments counter-propagating with respect to each other in their center-of-momentum reference frame. The considered configuration is shown in Fig.~\ref{fig1}(c) and reveals a strong density oscillation associated with interference of the order parameters (matter wave-packets) from the BEC pieces in the area of their overlapping. The existence of such a fringe structure of the density distribution has been directly observed as an effect of interference of two condensates in experiment \cite{Ketterle97}. The spatial phase matching condition, determined by internal relative motion of the fragments, crucially affects the scattering process. Indeed, the wave length of the oscillation is determined by the speed of relative motion and after accumulation of essential linear momentum from light can exceed a scale of the light wavelength. Then such a density grating should lead to strong Bragg diffraction and, as we show by our numerical simulations below, to significant enhancement of the backward scattering.

The process can be described by the order parameter of the following spatial profile
\begin{eqnarray}
\Xi(z)&=&\sqrt{2n_0}\cos \left( \frac{\pi z}{L}\right)\, \cos \left(\Delta q\,z\right)%
\nonumber\\
&=&\sqrt{\frac{n_0}{2}}\cos \left( \frac{\pi z}{L}\right)\,\mathrm{e}^{i\Delta q\,z}%
+\sqrt{\frac{n_0}{2}}\cos \left( \frac{\pi z}{L}\right)\,\mathrm{e}^{-i\Delta q\,z}%
\nonumber\\%
&\equiv& \Xi_{+}(z)+\Xi_{-}(z)%
\label{4.2}%
\end{eqnarray}
which is constructed as an ideal overlap of two matter wave-packets associated with the condensate fragments of identical shape and size counter-propagating with respect to each other with the relative linear momentum $2\hbar\Delta q$ per atom. Let us make a clarifying comment concerning the validity and consistency of the suggested profile as a solution of the time dependent Gross-Pitaevskii equation.

Both of the partial contributions $\Xi_{+}(z)$ and $\Xi_{-}(z)$ are representative solutions of the order parameter equation, for example, in the Thomas-Fermi approximation. That can be justified via transforming dynamical description of any of the wave-packets to that reference frame where the particular fragment is motionless and then we arrive to the configuration considered in the previous subsection. But the entire process of expansion and fragmentation of the condensate, modeled by (\ref{4.2}), can be imagined only after the BEC is released from the trap and it results from both the external disturbance and internal interaction processes. The superposed state (\ref{4.2}) can physically model the complicated dynamics of the condensate fragmentation once we ignore the weak non-ideality of the atomic gas in comparison with the kinetic energy associated with the relative motion of the fragments, see inequality (\ref{2.11}). This can be fulfilled for quite high relative speed with $\Delta\,q\gg 1/L$ and $2\hbar^2\Delta q^2/\mathrm{m_A}>\mu_c$. Then the factor "$\cos \left(\Delta q\,z\right)$" is a strongly oscillating function of $z$, which implies its averaging $\langle\cos^2\Delta q\,z\rangle\to 1/2$ in the normalization of the order parameter by a total number of particles. Then expansion (\ref{4.2}) corresponds to beginning of the splitting process of the released matter wave $\Xi(z)=\sqrt{n_0}\cos (\pi z/L)$, as shown in Fig.~\ref{fig1}(c), in two separated wave-packets $\Xi_{+}(z)$ and $\Xi_{-}(z)$ propagating in opposite directions.

\begin{figure}[tp]
{$\scalebox{0.8}{\includegraphics*{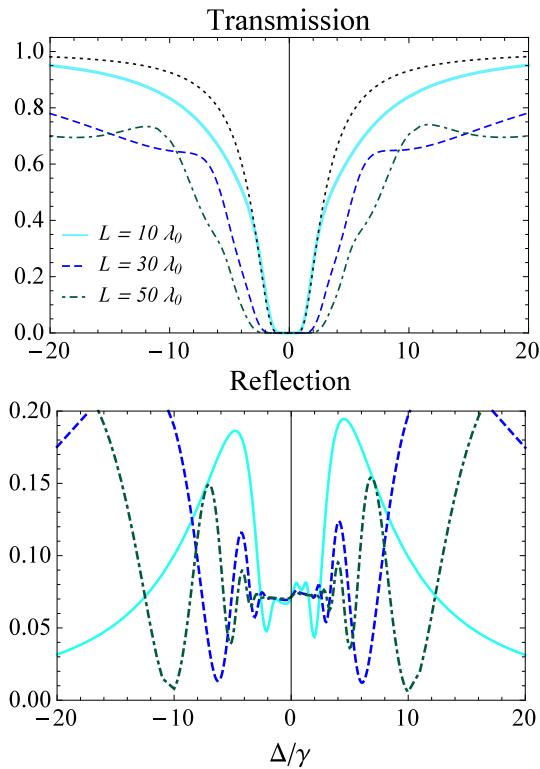}}$}
\caption{(Color online) Same as in Fig.~\ref{fig2} but for the density distribution parametrized by the order parameter $\Xi(z)=\sqrt{2n_0}\cos \left( \pi z/L\right)\, \cos \left(\Delta q\, z\right)$ with $2\Delta q=k_0$ for the geometry shown in Fig.~\ref{fig1}(c). Both the forward and backward scattering have a clear signature of the coherent enhancement due to the effect of the Bragg diffraction. In the upper panel the dotted curve indicates the reference transmission spectrum for $L=10\lambda_0$ with smoothed profile of the order parameter and corresponds to the configuration of a disordered atomic gas.}
\label{fig4}%
\end{figure}%

In Fig.~\ref{fig4} we show the spectra of transmission and reflection for the order parameter with the spatial profile given by Eq.~(\ref{4.2}). It is expected that for a classical disordered gas any internal motion of its macroscopic fragments with a rather slow relative speed would not modify the scattering process at all. As an example, such an expansion with a relative speed given by the recoil limit $\sim\hbar k_0/\mathrm{m_A}$ would induce only a negligible Doppler shift between the spectral outputs from both the fragments. But in the case of BEC such an internal motion dramatically modifies the scattering process. As pointed out above, the spatial modulation of the order parameter initiates a mechanism of the Bragg diffraction and scattering on the spatially oscillating density. As a consequence, this leads to strong enhancement of the backward scattering and it is manifestable in an abrupt structure of the transmission spectrum as well. The strongest scattering is observed for the modulation wave number $\Delta q=k_0$ when the condensate expands with the relative speed $v_0=2\hbar k_0/\mathrm{m_A}$. As follows from the dependencies of Fig.~\ref{fig4} this effect experiences as a broader spectral domain as the sample spatial scale is longer.

\begin{figure}[tp]
{$\scalebox{0.8}{\includegraphics*{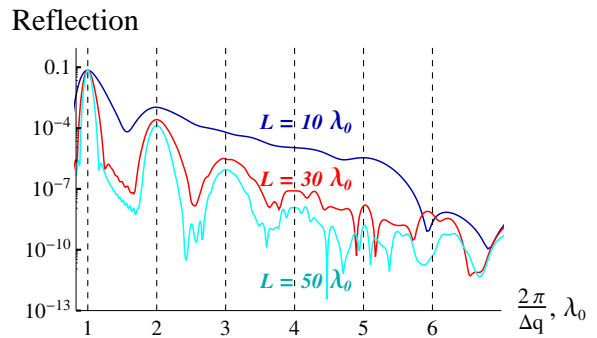}}$}
\caption{(Color online) Reflection coefficient for the order parameter of Fig.~\ref{fig4}, at the point of atomic resonance, plotted as a function of $2\pi/\Delta q$ for different sample depths $L$.}
\label{fig5}%
\end{figure}

In Fig.~\ref{fig5} we reproduce the dependence of the reflection coefficient as a function of $2\pi/\Delta q$ for different sample depths $L$. As can be seen from these graphs, the reflection always has a local maxima at the points $\Delta q=2\pi/\lambda_0,\,2\pi/2\lambda_0\ldots$. This is optimal condition for manifestation of the Bragg diffraction, which creates the oppositely propagating polariton wave via scattering of the impinging wave on periodic structure. As a consequence of the Bragg-type scattering an additional amount of linear momentum transfers to the condensate and enforces its fragmentation. So the Bragg diffraction also results in a certain optomechanical action on the system and accordingly leads to kinematic entanglement of the spatially structured BEC, see \cite{Hilliard}.

In our calculation model we can describe such an effect of optomechanical interface primary for the backward and forward scattering channels. Nevertheless, in experiment \cite{Schneble} the fragmentation was observed for the scattering directions orthogonal to the incident light along the major axis of an ellipsoid-shaped condensate sample. The observed effect had been associated in \cite{Schneble} with the Kapitza-Dirac phenomenon of the matter wave scattering on the spatial structure created by an electromagnetic wave. In this sense, we can point out that in the case of excitation of a BEC sample by an external light pulse, consisting of many photons, the entire dynamics apparently results from several physical processes, which includes internal interactions, disturbance of the matter-wave (order parameter) by external driving field and formation of the polariton structure by the optical excitation. Then the Bragg scattering reveals a coherent mechanism for rearranging of photon-type polariton waves (see Section \ref{Section_III_C}) propagating in different directions. The coherently scattered photons emerge the sample with indicating prior propagation directions of these waves.

\section{Conclusion}\label{Section_V}
\noindent In this paper we have developed a formalism of the microscopic quantum scattering theory directed towards ab-initio description of the elementary process of a single photon scattering from a quantum degenerate atomic gas. The gas exists in the BEC phase parametrized by the order parameter introduced in the framework of the Gross-Pitaevskii model. The main mathematical object of our calculational approach is the single particle Green's function (propagator) tracking the propagation of a specific polariton wave through the condensate. The polariton is created as a quantum superposed state between the photon and condensate.

The polariton propagation is disturbed by the process of incoherent scattering and its entire dynamics is described by the closed scattering equation for the complete polariton propagator as we derived. The crucial difference with the light propagation through a disordered and non-degenerate atomic gas is that in the considered case the atomic medium represents a coherent matter wave strongly rejecting its classical interpretation. The conventional vision of the macroscopic Maxwell description of the electromagnetic wave in a bulk medium seems insufficient and can be even incorrect in some situations.

To clarify the above point we have solved the derived scattering equations in a one-dimensional geometry and compared the result with predictions of the conventional macroscopic Maxwell theory for the disordered atomic gas of the same density and size as the BEC sample. For steady state conditions and uniform distribution of the order parameter we obtained identical results for the transmission and reflection spectra for both the approaches. Nevertheless we observe a significant difference once the BEC is fractured into a number of the interfering matter wave fragments. In the latter case the scattering process evolves towards conditions of Bragg diffraction, which strongly affects the process and can coherently redirect the propagating polariton wave in the backward or other directions associated with the condensate fragmentation.

\section*{Acknowledgements}

\noindent This work was supported by the Russian Foundation for Basic Research under Grant 15-02-01060.   We also acknowledge financial support by the National Science Foundation under Grant No. NSF-PHY-1606743.

\appendix
\section{Overview of the diagram approach}\label{Appendix_A}
\noindent Below we introduce basic elements of the diagram equations, which are constructed and discussed in the main text. We follow standard definitions and rules of the microscopic version of the Feynman diagram method, as described in Ref.~\cite{BerstLifshPitvsk}, but revise it for a non-relativistic dipole-type coupling of light and atoms, see Ref.~\cite{Kupriyanov17}. The expansion of the evolution operator (\ref{2.10}) in the Green's function (\ref{2.9}) generates the sequence of expectation values of the various operator products, which after a set of transpositions and with the aid of the Wick theorem can be regrouped to the results visualized by diagram images. The diagrams consist of the objects listed below.

The undisturbed causal-type electric field Green's function is defined via transposition of the field operators in any pair product from chronologically $T$-ordered to normally $N$-ordered form
\begin{eqnarray}
\lefteqn{\hspace{-0.5cm}iD_{\mu\mu'}^{(E)}(\mathbf{r},t;\mathbf{r}',t')}%
\nonumber\\%
&&\hspace{-1cm}=T\!\left[E^{(0)}_\mu\!(\mathbf{r};t)E^{(0)}_{\mu'}\!(\mathbf{r}';t')\right]\!-\!N\!\left[E^{(0)}_{\mu}\!(\mathbf{r};t)E^{(0)}_{\mu'}\!(\mathbf{r}';t')\right]%
\label{a.1}%
\end{eqnarray}
It can be linked with a fundamental object of quantum electrodynamics namely with the causal-type photon propagator
\begin{equation}
D_{\mu\mu'}^{(E)}(\mathbf{r},t;\mathbf{r}',t')=\left.\frac{1}{c^2}\frac{\partial^2}{\partial t\partial t'}%
D_{\mu\mu'}^{(c)}(\mathbf{r},t;\mathbf{r}',t')\right|_{\scriptsize{\begin{array}{c}\mathbf{r}\neq\mathbf{r}'\\\mathrm{or}\\ t\neq t'\end{array}}}%
\label{a.2}%
\end{equation}
where we follow gradient invariance of the theory and fix the propagator by a vanishing scalar potential such that $\mu,\,\mu'=x,y,z$. With simplifying notation for each argument  $\mu,\mathbf{r},t\to x$ and $\mu',\mathbf{r}',t'\to x'$ the electric field Green's function is imaged by a wavy line
\begin{equation}
iD^{(E)}(x,x')\Leftrightarrow\begin{array}[c]{c}\scalebox{1.0}{\includegraphics*{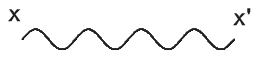}}\end{array}%
\label{a.3}%
\end{equation}
where the ending indices are often omitted in graph equations. This function depends only on the difference of its spatial and time arguments and its Fourier image is given by
\begin{eqnarray}
\lefteqn{D^{(E)}_{\mu\mu'}(\mathbf{k},\omega)}%
\nonumber\\%
&&=\!\int\!d^3\!R\int^{\infty}_{-\infty}\!\!d\tau\,%
\mathrm{e}^{i\omega\tau-i\mathbf{k}\cdot\mathbf{R}}\!\left.D_{\mu\mu'}^{(E)}(\mathbf{R},\tau)\right|_{\scriptsize{\begin{array}{c}\mathbf{R}=\mathbf{r}-\mathbf{r}'\\ \tau=t-t'\end{array}}}%
\nonumber\\%
&&=\!\frac{4\pi\hbar\omega^2}{\omega^2-\omega^2_k+i0}\left[\delta_{\mu\mu'}-c^2\,\frac{k_{\mu'}k_{\mu}}{\omega^2}\right]%
\label{a.4}
\end{eqnarray}
where $\omega_k=ck$.

The electric field Green's function is expressed via solution of the microscopic Maxwell equations with a point-like dipole source and for $\omega>0$ coincides with positive-frequency component of the retarded-type fundamental solution of these equations $D^{(R)}_{\mu'\mu}(\mathbf{k},\omega)$
\begin{equation}
\left.D^{(E)}_{\mu\mu'}(\mathbf{k},\omega)\right|_{\omega>0}=\frac{\omega^2}{c^2}\left.D^{(R)}_{\mu\mu'}(\mathbf{k},\omega)\right|_{\omega>0}
\label{a.5}%
\end{equation}
The positive frequency domain is only important in the RWA approach and in this approximation it is convenient to add an arrow in the diagram (\ref{a.3}) for indicating creation and annihilation events of a virtual photon at the edging points of the wavy line.

The undisturbed atomic Greens's function is defined via transposition of the atomic operators in any pair product from chronologically ordered to normally ordered form. For operators of the excited state this reads
\begin{eqnarray}
\lefteqn{\hspace{-0.5cm}iG_{nn'}^{(0)}(\mathbf{r},t;\mathbf{r}',t')}%
\nonumber\\%
&&\hspace{-0.8cm}=T\!\left[\Psi^{(0)}_n(\mathbf{r};t)\Psi^{(0)\dagger}_{n'}(\mathbf{r}';t')\right]\!-\!\Psi^{(0)\dagger}_{n'}(\mathbf{r}';t')\Psi^{(0)}_n(\mathbf{r};t)%
\label{a.6}%
\end{eqnarray}
and similarly with replacement $n,n'\to m\!=\!m'\!=0$ for operators of the ground state. With simplifying notation for each argument  $n,\mathbf{r},t\to x$ and $n',\mathbf{r}',t'\to x'$ the atomic Green's function is imaged by an arrowed straight line
\begin{equation}
iG^{(0)}(x,x')\Leftrightarrow\ \ \begin{array}[c]{c}\scalebox{1.0}{\includegraphics*{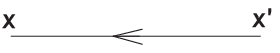}}\end{array}%
\label{a.7}%
\end{equation}
where the ending indices are often omitted in graph equations. This function also depends only on the difference between its spatial and time arguments and its Fourier image is given by
\begin{eqnarray}
\lefteqn{G^{(0)}_{nn'}(\mathbf{p},E)}%
\nonumber\\%
&&\hspace{-0.5cm}=\!\int\!d^3\!R\int^{\infty}_{-\infty}\!\!d\tau\,%
\mathrm{e}^{\frac{i}{\hbar}E\tau-\frac{i}{\hbar}\mathbf{p}\cdot\mathbf{R}}\!\left.G_{nn'}^{(0)}(\mathbf{R},\tau)\right|_{\scriptsize{\begin{array}{c}\mathbf{R}=\mathbf{r}-\mathbf{r}'\\ \tau=t-t'\end{array}}}%
\nonumber\\%
&=&\delta_{nn'}\,\frac{\hbar}{E-p^2/{2\mathrm{m_A}}-E_n+i0}%
\label{a.8}
\end{eqnarray}
where $\mathrm{m}_A$ is the atomic mass and the internal atomic state is assumed to be degenerate such that $E_n=\mathrm{const}_n$.

The atomic Green's function is expressed by the fundamental solution (atomic propagator) of the Schr\"{o}dinger equation for a free atom which describes propagation of an atomic wave initially localized in a certain spatial point. As follows from (\ref{a.6}) this function vanishes if $t<t'$ such that the causal-type atomic propagator is identical to the retarded-type propagator.

There are different diagram vertices indicating optical interactions of different types. If a virtual photon interacts with an atom, which is also presented as a virtual object in a diagram, then in the RWA we associate the process with the following two vertexes
\begin{eqnarray}
\frac{i}{\hbar}\,d^{\mu}_{nm}\ &\Leftrightarrow&\ \ \ \raisebox{-0.3 cm}{\scalebox{0.7}{\includegraphics*{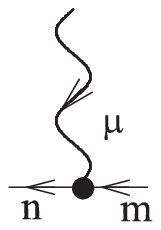}}}%
\nonumber\\%
\frac{i}{\hbar}\,d^{\mu}_{mn}\ &\Leftrightarrow&\ \ \ \raisebox{-0.3 cm}{\scalebox{0.7}{\includegraphics*{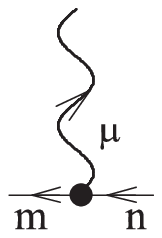}}}%
\label{a.9}
\end{eqnarray}
If a similar process is developing with condensate particles we associate it with the vertexes of another type
\begin{eqnarray}
\frac{i}{\hbar}\,d^{\mu}_{nm}\,\Xi(\mathbf{r})\,\mathrm{e}^{-\frac{i}{\hbar}\epsilon_0t}\ &\Leftrightarrow&\ \ \raisebox{-0.3 cm}{\scalebox{0.45}{\includegraphics*{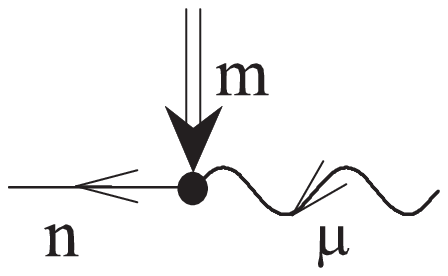}}}%
\nonumber\\%
\frac{i}{\hbar}\,d^{\mu}_{mn}\,\Xi^{\ast}(\mathbf{r})\,\mathrm{e}^{+\frac{i}{\hbar}\epsilon_0t}\ &\Leftrightarrow&\ \ \raisebox{-0.3 cm}{\scalebox{0.45}{\includegraphics*{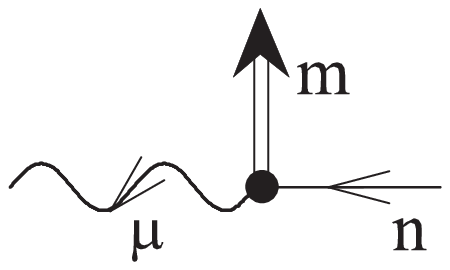}}}%
\label{a.10}
\end{eqnarray}
which describe either excitation of an atom from the condensate phase (upper diagram) or its recovering in the condensate phase (lower diagram). The detailed specification of vertexes is usually unnecessary and often omitted if it does not confuse interpretation of the diagram.

In the original representation each vertex corresponds to the integral over respective spatial and time variables and each contributing line is decoded in accordance with (\ref{a.3}) and (\ref{a.7}). In the stationary and homogeneous conditions after Fourier transform, the external lines are decoded in accordance with (\ref{a.4}) and (\ref{a.8}) but internal lines, when they shape a loop, contribute as convolution-type integrals over reciprocal variables such as energy (frequency) and momentum (wave vector), with conserving total energy and momentum transporting by the diagram. For more details we readdress reader to Refs.\cite{BerstLifshPitvsk,Kupriyanov17}.

\section{One-dimensional scattering}\label{Appendix_B}
\noindent Consider the scattering problem for a slab geometry of an atomic medium, homogeneous and infinite in the plane transverse to the wave vector of the incident photon. In this case the $T$-matrix element, given by Eq.~(\ref{2.6}) and selected for either forward or backward elastic scattering channels, is given by
\begin{eqnarray}
\lefteqn{\hspace{-0.5cm}T_{i'i}(E)=\frac{2\pi\omega}{{\cal L}}\iint\,dz'\,dz\sum\limits_{n',n}{(\mathbf{d}\cdot\mathbf{e})}^*_{n'0}{(\mathbf{d}\cdot\mathbf{e})}_{0n}}%
\nonumber\\%
&&\hspace{-0.5cm}\times{\mathbf e}^{-ik'z'+ikz}\;\Xi^*(z')\,\Xi(z)\,G_{n'n}(z',z;E-E_0^{N-1})%
\label{b.1}%
\end{eqnarray}
where the output frequency and polarization are unchanged such that $\omega'=\omega$ and $\mathbf{e}'=\mathbf{e}$, and we redefined $f=i'$ with emphasizing the physical equivalence of initial and final states in one dimensional scattering process. All the integrands are considered as functions of longitudinal coordinates $z,z'$ and the polariton propagator is proportional to a $\delta$-function of transverse coordinates $x,y$ and $x',y'$, see Eq.~(\ref{3.19}). The integral evaluated in the transverse plane over variables $dxdy$ and $dx'dy'$ cancels out the area scale ${\cal L}_x{\cal L}_y$ in the normalization volume ${\cal V}={\cal L}_x{\cal L}_y{\cal L}_z$ and we denoted ${\cal L}_z={\cal L}$.

Let us express the $S$-matrix components via $T$-matrix
\begin{eqnarray}
S_{i'i}&=&\delta_{i'i}-i\,\frac{{\cal L}}{\hbar c}\,T_{i'i}(E_i+i0)%
\label{b.2}
\end{eqnarray}
In a one-dimensional geometry for non-degenerate ground state of the degenerate quantum gas the light scattering can be described by coefficients of transmission $\mathrm{T}(\omega)$, reflection $\mathrm{R}(\omega)$ and losses $\mathrm{L}(\omega)$, which are subsequently given by
\begin{eqnarray}
\mathrm{T}(\omega)&=&\left.{\left|S_{i'i}\right|^2}\right|_{k'=k>0}%
\nonumber\\%
\nonumber\\%
\mathrm{R}(\omega)&=&\left.{\left|S_{i'i}\right|^2}\right|_{k'=-k<0}%
\nonumber\\%
\nonumber\\%
\mathrm{L}(\omega)&=&1-\mathrm{T}(\omega)-\mathrm{R}(\omega)%
\label{b.3}%
\end{eqnarray}
and can be found via solution of the simplified equations (\ref{3.19})-(\ref{3.21}) as we show below.

Consider the example of the slab with the order parameter $\Xi(z)=\sqrt{n_0}=\mathrm{const}_z$ inside the medium. In this case the integral equation (\ref{3.19}) can be transformed to the set of algebraic equations via spatial Fourier transform with periodic boundary conditions on the sample bounds. The azimuthal symmetry justifies the diagonal structure of the polariton propagator
\begin{equation}
G_{nn'}(z,z';E)=\delta_{nn'}\,G(z,z';E)%
\label{b.4}%
\end{equation}
Then, with the assumption that the origin of the coordinate frame is located in the middle point and $z\in(-L/2,L/2)$, where $L$ is the sample length, it can be expanded as
\begin{eqnarray}
G_{ss'}(E)&=&\frac{1}{L}\,\iint\limits_{-L/2}\limits^{L/2}dzdz'\mathrm{e}^{-ik_sz+ik_{s'}z'}G(z,z';E)%
\nonumber\\%
G(z,z';E)&=&\frac{1}{L}\,\sum_{s,s'}\mathrm{e}^{ik_sz-ik_{s'}z'}G_{ss'}(E)%
\label{b.5}%
\end{eqnarray}
where $k_s=2\pi s/L$ and $k_{s'}=2\pi s'/L$ with $s,s'=0,\pm 1,\pm 2,\ldots$. The Green's function (\ref{b.4}) contributes to the transmission amplitude (\ref{b.2}) at a specific energy argument $E-E_0^{N-1}=E_i-E_0^{N-1}=\hbar\omega+E_0^{N}-E_0^{N-1}=\hbar\omega+\varepsilon_0$ and we denote
\begin{equation}
\left.{G_{ss'}(E)}\right|_{E=\hbar\omega+\varepsilon_0}\equiv G_{ss'}(\omega)%
\label{b.6}%
\end{equation}
and consider the Fourier components as functions of the frequency of the incident photon. Then the $S$-matrix elements (\ref{b.2}) can be expressed as follows
\begin{eqnarray}
\lefteqn{S_{i'i}=\delta_{i'i} - \frac{8\pi i\omega}{L\hbar c}\,n_0d_0^2\,\sum_{s',s}}
\nonumber\\%
&&\times\frac{\sin\left(k'-k_{s'}\right)\frac{L}{2}}{k'-k_{s'}}\;\frac{\sin\left(k-k_s\right)\frac{L}{2}}{k-k_s}\;G_{s's}(\omega)
\label{b.7}%
\end{eqnarray}
where $k=\omega/c$ and $k'=\pm\omega/c$.

With substituting (\ref{b.4}) and applying transforms (\ref{b.5}) to Eq.~(\ref{3.19}), considered in a one-dimensional configuration, we arrive at the following system of algebraic equations
\begin{eqnarray}
\lefteqn{\hspace{-2cm}\left[\omega-\tilde{\omega}_0+\frac{\hbar\,k_s^2}{2\mathrm{m_A}}+\frac{4\pi}{3\hbar}\,n_0\,d_0^2+\frac{i}{2}\sqrt{\epsilon\left(\omega\right)}\;\gamma\right]G_{ss'}(\omega)}%
\nonumber\\%
&&\hspace{-1cm}-\sum_{s''}\Sigma_{ss''}^{(c)}(\omega)G_{s''s'}(\omega)%
=\delta_{ss'}%
\label{b.8}%
\end{eqnarray}
where $\tilde{\omega}_0=(E_n-\varepsilon_0)/\hbar=(E_n-E_0-\mu_c)/\hbar\equiv\omega_0-\mu_c/\hbar$ with same $E_n$ for all the upper state Zeeman sublevels. We approximated $\epsilon(\omega+\mu_c/\hbar)\approx\epsilon(\omega)$, see Eq.~(\ref{3.17}) and related comment.

The matrix of the self-energy part is given by
\begin{eqnarray}
\lefteqn{\Sigma_{ss}^{(c)}(\omega)=\frac{4\pi}{\hbar}n_0d_0^2\,\frac{\omega^2}{\omega^2-c^2\,k_s^2}}%
\nonumber\\%
&&-\frac{4\pi i}{\hbar}n_0d_0^2\,%
\frac{\omega}{cL}\frac{\displaystyle\frac{\omega^2}{c^2}+k_s^2}{\left(\displaystyle\frac{\omega^2}{c^2}-k_s^2\right)^2}\left[1-\exp\left(i\frac{\omega}{c}L\right)\right]%
\label{b.9}
\end{eqnarray}
for $s''=s$ and
\begin{eqnarray}
\Sigma_{ss''}^{(c)}(\omega)\!\!&=&\!\!-(-)^{s-s''}\frac{4\pi i}{\hbar}n_0d_0^2\,%
\frac{\omega}{cL}\frac{\displaystyle\frac{\omega^2}{c^2}+k_sk_{s''}}{\left(\displaystyle\frac{\omega^2}{c^2}-k_s^2\right)\!\!\left(\displaystyle\frac{\omega^2}{c^2}-k_{s''}^2\right)}%
\nonumber\\%
&\times&\left[1-\exp\left(i\frac{\omega}{c}L\right)\right]%
\label{b.10}%
\end{eqnarray}
for $s''\neq s$. For a sample of infinite length $L\to\infty$ Eqs.~(\ref{b.6}), (\ref{b.8})-(\ref{b.10}) reproduce the transverse component of the polariton propagator in an infinite and uniform medium, see Eq.~(\ref{3.23}), and in this case the scattering process manifests itself mainly via the incoherent channels.

For the sample of finite length the system (\ref{b.8}) consists of an infinite number of equations. Nevertheless it can be numerically solved with cutoff by a limited number of the involved equations. With increasing of this number the iterative process becomes internally converging and approaching the exact solution. The performed calculation scheme can be straightforwardly generalized if the order parameter is non-uniform and described by trigonometric functions such as $\Xi(z)\sim\cos (\pi z/L)$, $\Xi(z)\sim\mathrm{e}^{i\kappa_1z}\cos (\pi z/L)+\mathrm{e}^{i\kappa_2z}\cos (\pi z/L)$ etc., which we have considered in our numerical simulations.

\end{document}